\documentstyle[emulateapj]{article}

\newcommand{\etal}{et~al.}

\newcommand{\gtsim}{\mathrel{\hbox{\rlap{\lower.55ex \hbox {$\sim$}}
                   \kern-.3em \raise.4ex \hbox{$>$}}}}

\newcommand{\ltsim}{\mathrel{\hbox{\rlap{\lower.55ex \hbox {$\sim$}}
                   \kern-.3em \raise.4ex \hbox{$<$}}}}

\newcommand{\msun}{\mbox{\rm\,M$_{\odot}$}}

\newcommand{\refto}{\cite}

\begin{document}
\title{Discovery of a luminous white dwarf\\
       in a young star
cluster in the Large Magellanic Cloud}
\author{Rebecca A.W. Elson\altaffilmark{1},
        Steinn Sigurdsson\altaffilmark{2}
Jarrod Hurley,\\
Melvyn B. Davies\altaffilmark{3}, and Gerard F. Gilmore\\
        Institute of Astronomy, Madingley Road, Cambridge CB3 0HA, UK}

\altaffiltext{1}{corresponding author; E-mail: elson@ast.cam.ac.uk
}

\altaffiltext{2}{EU Marie Curie Fellow; E-mail: steinn@ast.cam.ac.uk
}

\altaffiltext{3}{Royal Society URF
}

\date{\underline{submitted to {\sl Astrophysical Journal\/}, 6 Feb 1998}
}


\begin{abstract}

We have identified a candidate $\sim 1-2 \times 10^5$ year old 
luminous white dwarf in NGC 1818, a young 
star cluster in the Large Magellanic Cloud. This 
discovery strongly constrains
the boundary mass $M_c$
at which stars stop forming neutron stars and start forming
white dwarfs, to $M_c \gtsim 7.6 \msun$.


\end{abstract}

\section{Introduction}

Stars of mass comparable to that of the sun evolve to form white dwarfs
while, above some critical mass, $M_c$,
stars detonate as type II supernovae instead,
leaving neutron stars as remnants.
The determination of $M_c$
is important for understanding stellar evolution.
It also has implications for the chemical evolution of galaxies,
in that it affects the rate of enrichment of the interstellar medium
and the total population of pulsars.
The relationship between the mass of a white dwarf and that of its
progenitor, as a function of the metallicity of the progenitor,
is also important for understanding stellar evolution in the early 
universe (\refto{Jeff,vdB}).
Predictions of $M_c$ range from $6-10\, \msun$ 
depending on the details of
models and on the metallicity of the star (\refto{Weid,Jeff,gar97,Hur}). 
Almost no observational constraints on $M_c$ are available 
except from one Galactic open
cluster in which the most luminous of four white dwarfs implies
$M_c > 5.7 \, \msun$ (\refto{Koe}). Statistical extrapolations of these data
suggest $M_c \sim 8 \, \msun$ but with large formal uncertainty (\refto{Jeff}).

The young rich star clusters in the Large Magellanic Cloud (LMC) are
particularly well suited for studies of the evolution of
intermediate mass stars.  They typically contain
an order of magnitude more stars than Galactic open clusters, and 
membership determination is not generally problematic as it is with
Galactic clusters, which are often superposed against a dense curtain of
disk stars.
The clusters with age $\sim 2-4 \times 10^7$ yr have main-sequence
turnoff masses $\sim 7.5 - 9.5 \msun$, (\refto{Will})
interestingly
close to the theoretical range of values for $M_c$.

Models show that turnoff stars of this
mass spend $\sim2\times 10^6$ yr on the red 
giant branch (\refto{Hur,Scha92})
before ejecting their hydrogen envelope and
leaving  rapidly fading white dwarfs
(assuming they in fact have masses $< M_c$).
Rich LMC clusters typically have $\sim 10-20$ red giant branch stars, so we would
expect the youngest remnant
star to be $\sim 1-2 \times 10^5$ years old.
The models of Wood give
luminosities for 
white dwarfs of this mass and age of $30-100 L_\odot$,
with temperatures of $\sim $100,000 K (\refto{Woo92,Woo2,dan}). Cooling is 
dominated by neutrino emission, and the remnants fade 
proportionally to  $\sim t^{-2.5} $ or faster, depending on model details.

Adopting a distance modulus for the LMC of  
$(V-M_V)=18.5$ (\refto{Pan91}) would 
imply apparent magnitudes for these young white dwarfs 
of $V\sim 18-19.5$. 
They would be visible even from the ground, although
because of crowding, Hubble Space Telescope 
(HST) observations would be required for accurate photometry.
The white dwarfs  would be
distinguished from comparably bright main-sequence 
stars by their extreme blueness. 
To investigate the possibility of finding young white dwarfs in
LMC clusters, we analysed HST archive images
of NGC~1818.
This cluster has  mass $\sim 2.8 \times 10^4 \msun$,
core radius $r_c=2.0$ pc, half-mass radius $r_h \sim 14$ pc (\refto{Els87}), and
age $\sim 2-4 \times 10^7$ yr (\refto{Will}).
There are currently $\sim 16$ stars in
the red giant phase.

\section {Observations}

The images of NGC 1818 were obtained with the Wide Field and Planetary Camera
(WFPC2) on 1995 December 8, with the F336W ($\equiv U_{336}$),
F555W ($\equiv V_{555}$), and F814W ($\equiv I_{814}$) filters.  
Total exposure times
are 960, 880 and 1290 seconds respectively (\refto{Hun97}).
The images in each filter were coadded with a median
filter to eliminate cosmic rays.  

DAOFIND was used to  detect objects
$4 \sigma$ above the background.  Point-spread function (PSF) fitting was 
used to help eliminate spurious detections: these include primarily structure in
the PSF, particularly  around saturated stars, and  bright pixels
along diffraction spikes.  The final photometry was performed using
an aperture with radius 2 pixels, and aperture corrections and zero
points were applied (\refto{Hola,Holb,Els98}).  A value of
$E(B-V)=0.05$ was adopted for the reddening.

A color-magnitude diagram (CMD) in $V_{555}$ vs $(U_{336}-V_{555})$ 
for the Planetary Camera (PC) revealed a
prominent sequence of binary stars, and we explore the implications
for the binary population of NGC 1818 elsewhere (\refto{Els98}).  
CMDs for the
three WFC chips combined are shown in Figs. 1a and b.
These cover a radial range
from $\sim 2$ core radii, out to about half way to the edge of the cluster.
Stars with $V_{555} < 17.5$ are saturated;  the main-sequence turnoff is
at $V_{555} \approx 14$.
(We note the presence of an apparent gap in the main-sequence
at $V_{555}\sim 20.3$ which may be due to a
possible jump in stellar magnitudes around $\sim 1.5 - 2 \msun$
due to the onset of convective overshooting (\refto{Hur}).)

Models suggest that white dwarfs should have $(U-V)_0 \sim -1.5$ and 
$(V-I)_0 \sim -0.4$ (\refto{Woo2,Che93});
these values are roughly independent of age
and metallicity over the range of interest.
For white dwarfs 
the model colors are equivalent to $(V_{555}-I_{814})=-0.44$
and $(U_{336}-V_{555}) = -1.85$ 
(\refto{Holb} Fig. 11).

The boxes 
drawn in Figs. 1a and b delimit plausible ranges of
colors and magnitudes for a young white dwarf.  
We inspected visually all detections within these boxes,
and in a similar range of magnitudes and
colors for the CMDs for the PC chip.
All but one of the objects 
turned out to be spurious detections as described above.
One, however, turned out to be unambiguously stellar.  It
is indicated with a filled circle in the CMDs, and  is circled on the
image of the cluster shown in Figure 2.

Could the photometry, or identification as a white dwarf, be in error?
Poisson error bars are included in Fig. 1b, and indicate that
the extreme blueness of this object cannot be due to random photometric
errors.  Nor is it
likely that the colors are biased bluewards
by a residual cosmic ray in the F336W image:  six independent images
were coadded, so a cosmic ray would have had to hit the same pixel in 
several of them.  Goodness-of-fit parameters (DAOPHOT's `sharpness'
parameter and $\chi^2$) derived from fitting
a PSF to this object 
are within the range expected for a star:  there is no
evidence that it is strongly peaked, as a cosmic ray would
be, or resolved, as a background galaxy would be. (A 
background galaxy this bright would be easily identifiable in a WFPC2 image.)

The object is extremely
unlikely to be a quasar:  quasars of this magnitude have surface density
$\sim 1$ per square degree (\refto{Co,Boy}), so the probability of finding one
in our field is $\sim 10^{-3}$.  It is even more unlikely to be a foreground 
star in the halo of our Galaxy:  the number of Galactic stars of this 
magnitude expected in a WFPC2 field of view at high galactic latitude is
less than one (\refto{San}).
At a distance of, say, 10 kpc, a Galactic white dwarf
would have absolute magnitude
$M_V \sim -3.5$, which is far too bright to  
be an old white dwarf. 
On the other hand, given the number of stars on the giant branch,
we would expect to find about one young white dwarf if 
the turnoff mass in NGC 1818 is indeed less than $M_c$.
Nor should we be surprised to find it outside the
core of the cluster:  more than half the red giants in the cluster
are outside the core, and four are further from the cluster center
than our white dwarf candidate. 

If we see one white dwarf with $V \sim 18.5$, would we expect to
see many more fainter ones?  This depends on the cluster mass function 
and on the difference between the turnoff mass and  $M_c$.
Given the number of red giants in the cluster, and assuming 
that $M_{turnoff} \approx M_c$, 
we might expect there to be $2 \pm 2$
other white dwarfs with magnitude up to $\sim 4$ mag fainter
than our candidate (ie. with $18.5 < V < 22.5$).
Given that sample of stars in the PC image is incomplete 
due to the presence of many saturated stars, and that
the WFC chips cover just over 50\% of the outer parts of the
cluster, the expected numbers are consistent with our
observations.
If $M_c$ is significantly greater than the turnoff mass of the cluster 
then stars with slightly smaller masses than the current turnoff mass
will have evolved to form white dwarfs which by now will have cooled
and faded.  We might therefore expect there to
be 4--8 other white dwarfs at fainter magnitudes still; 
the exact number expected is dominated by small number statistics when the
mass function is extrapolated.
Due to the rapid fading, we would expect these to
be $\sim 5$ magnitudes fainter (ie. $V \sim 23.5$) and  
they would be difficult or impossible to detect 
in the currently available  WFPC2 images.
The luminosity function of white dwarfs expected in young LMC
clusters is discussed further in \refto{Els98}.
Deeper images of NGC 1818 are scheduled during HST Cycle 7
(Project 7307) to look for
a sequence of fainter, older white dwarfs.

\section {Discussion}

What constraints on $M_c$ can we infer from the presence of a young
white dwarf in NGC 1818?  $M_c$ must be greater
than the turnoff mass, so the main task is to determine an accurate
value for this.  The main uncertainties are not in determining the
magnitude of the turnoff, but in converting this magnitude to a 
mass.  This requires a knowledge of both the metallicity and of
whether stellar evolution models with
or without convective overshooting are more appropriate. 
The metallicity of NGC 1818 has been determined from spectra
of two stars to be  [Fe/H]$= -0.8$ (\refto{Will}).
This value is, however, lower than the expected value for the young
population in the LMC (\refto{Olsz}), which is thought to be more like $-0.2$.
The membership of one of the stars is doubtful, and 
$-0.8$ may in fact be too low.
Two models with convective overshooting, and metallicities [Fe/H]$= -1.3$
and $-0.4$  give an
age for NGC 1818 of $4 \times 10^7$ yr, and a turnoff mass 7.6$\pm 0.1 \msun$.
Two models without overshooting give an age $2.2 \pm 0.2 \times 10^7$ yr,
and  a turnoff mass 9.0$\pm 0.5 \msun$ (\refto{Will,Scha92}).
From our own stellar evoultionary 
models, which include convective overshooting, we predict  
$M_c=7.0$ for [Fe/H]$= -0.8$ and $M_c=7.7 \msun$ for [Fe/H]$= -0.3$.
The convective overshoot parameter for these models is $\Lambda_c = 0.28$
for $M \sim 7.5 \msun$,
which is close to but slightly higher than the models of Schaller \etal
used by Will \etal  ~Other groups use different values for $\Lambda_c$,
and a full discussion of the effect of the choice of this parameter
on our results is deferred to a future paper (\refto{Els98}). 
Our observations imply $M_c \gtsim 7.6 \msun$, and perhaps
$M_c \gtsim 9.0 \msun$.  
A better determination of the metallicity of NGC 1818
would help constrain the value of $M_c$ further, as would deeper images in which
any older white dwarfs would be visible.

Another uncertainty affecting observational constraints on
$M_c$ is the possiblity of
an age spread among the stars near the 
main-sequence turnoff.
For example, 
it is possible that the lower mass stars formed first, by a sufficient 
margin that some stars slightly below top of the main-sequence
are evolving on to the red giant branch at the same time as more massive stars.
The most massive stars have main-sequence lifetimes about 5 Myr, 
so the total age spread is at most  a few Myr.
Such a spread in age would mean that stars evolving on to the red giant branch
could have masses $\ltsim 0.3 \msun$ less than the measured turnoff 
mass which would imply $M_c \gtsim 7.3 \msun$. 
Measuring such an age spreads requires accurate photometry
for stars just below the turnoff, and we will be acquiring this
during Cycle 7.  
Another source of uncertainty is the possibility that the
object is a member of
a mass--transfer binary, but this is unlikely and would require it
now be a tight white dwarf -- neutron star  binary.

Because young white dwarfs are bright and very blue, detecting candidates
even in 
ground based data is not difficult.  They would be expected to be present
in any star forming region containing significant numbers of stars
with masses near $M_c$.  For example, a recent CMD of an association in
the Small Magellanic Cloud containing stars with ages $10-60$ Myr
contains three stars whose colors and magnitudes are consistent with
those of young white dwarfs (\refto{dem}).

Future HST observations of NGC 1818 
and other young LMC clusters, to 
determine the cooling sequence, ages and masses of the
white dwarf population in these clusters, should 
allow us to determine $M_c$ more precisely,
and possibly for a range of metallicities.

\section{Summary}

We have identified a candidate luminous white dwarf in the young star cluster
NGC 1818 in the LMC.  The object is $\sim 35 $ arcsec from the cluster
center (about $4.5 r_c$ and $0.6 r_h$).  It has coordinates
5:04:13.8, $-$66:26:33.4 (J2000).  In the HST passbands it has
$V_{555}=18.43$, $(V_{555}-I_{814})=-0.26$ and $(U_{336}-V_{555})=-1.67$.
In the Johnson-Cousins system this corresponds to $V=18.44$, 
$(V-I)=-0.25$ and $(U-V)=-1.32$.  These values are corrected for
reddening assuming $E(B-V)=0.05$.  Posisson errors are $\pm 0.03$ for
$V$ and $\pm 0.04$ for the colors.  These do not include uncertainties
in the transformation to the Johnson-Cousin system.  
The temperature is probably $\gtsim 20,000$ K but is poorly constrained
by the $(U-V)$ color. 
With the adopted distance modulus of 18.5, the object has absolute
magnitude $M_V=-0.06$.

If this object is indeed a white dwarf, then
its mass is probably $ 1.1-1.3 \msun$ (\refto{Weid}). The composition of
white dwarfs formed from
high mass progenitors is expected to be Oxygen--Neon--Magnesium,
but may be Carbon/Oxygen. A spectroscopic determination of its
composition is a priority.
If spectroscopic followup observations confirm the identity of
the candidate star as a luminous young white dwarf, we have
strongly constrained the critical mass at which stars stop
evolving to type II supernovae to $M_c \gtsim 7.6 \msun$.

\smallskip
\noindent
{\bf Note:}
A preliminary spectrum of the white dwarf candidate
was obtained at the AAT 5 March 1998.  The 
spectrum rules out the possibility that the object is 
a quasar.  The 
velocity is indistinguishable from that of two other cluster members
so it is also very unlikely to be a foreground object.
Detailed modelling
of the spectrum is 
currently in progress.

\medskip
\noindent
{\bf Acknowledgements:\/}
This research was supported in part by a
PPARC rolling grant. SS acknowledges the support of the European
Union through a Marie Curie Individual Fellowship.
MBD gratefully acknowledges the support of the Royal Society through a URF.
Funding for JH was provided by a grant from the Cambridge Commonwealth
Trust, and from Trinity College.
We would like to thank Brian Boyle, Matt Burleigh, Helen Johnston and
Ray Stathakis for obtaining a spectrum of the candidate white
dwarf.



\newcommand{\journ}[4]{ {\sl #1\/} {\bf #2}, #3 (#4).}

\newpage

\noindent

\begin{figure}[h]
   \caption{Colour-magnitude diagrams for stars in NGC 1818
in the three WFC chips.  Magnitudes are in the HST pass-bands and have
been dereddened assuming $E(B-V)=0.05$.  Stars above $V_{555} \sim 17.5$
are saturated.  
The boxes indicate the range of colors and magnitudes
where we might expect to find white dwarfs.  All points in these boxes
are spurious detections except the filled circle which is a candidate
young white dwarf.  Poisson errors in the $(U_{336}-V_{555})$ colors
are shown. 
Much of the scatter redwards from the
main-sequence  at $V_{555} < 21$ is due to the presence of binary stars.}
   \label{fi:cmd}
\end{figure}

\begin{figure}[h]
   \caption{WFPC2 image of the inner regions of NGC 1818 in the  F555W
passband.
The field is 130 arcseconds on a side. The core radius of the cluster
is 8 arcsec.  The candidate white dwarf is circled. 
(The diameter of the circle is 7 arcsec.)}
   \label{fi:wdpic}
\end{figure}

\newpage

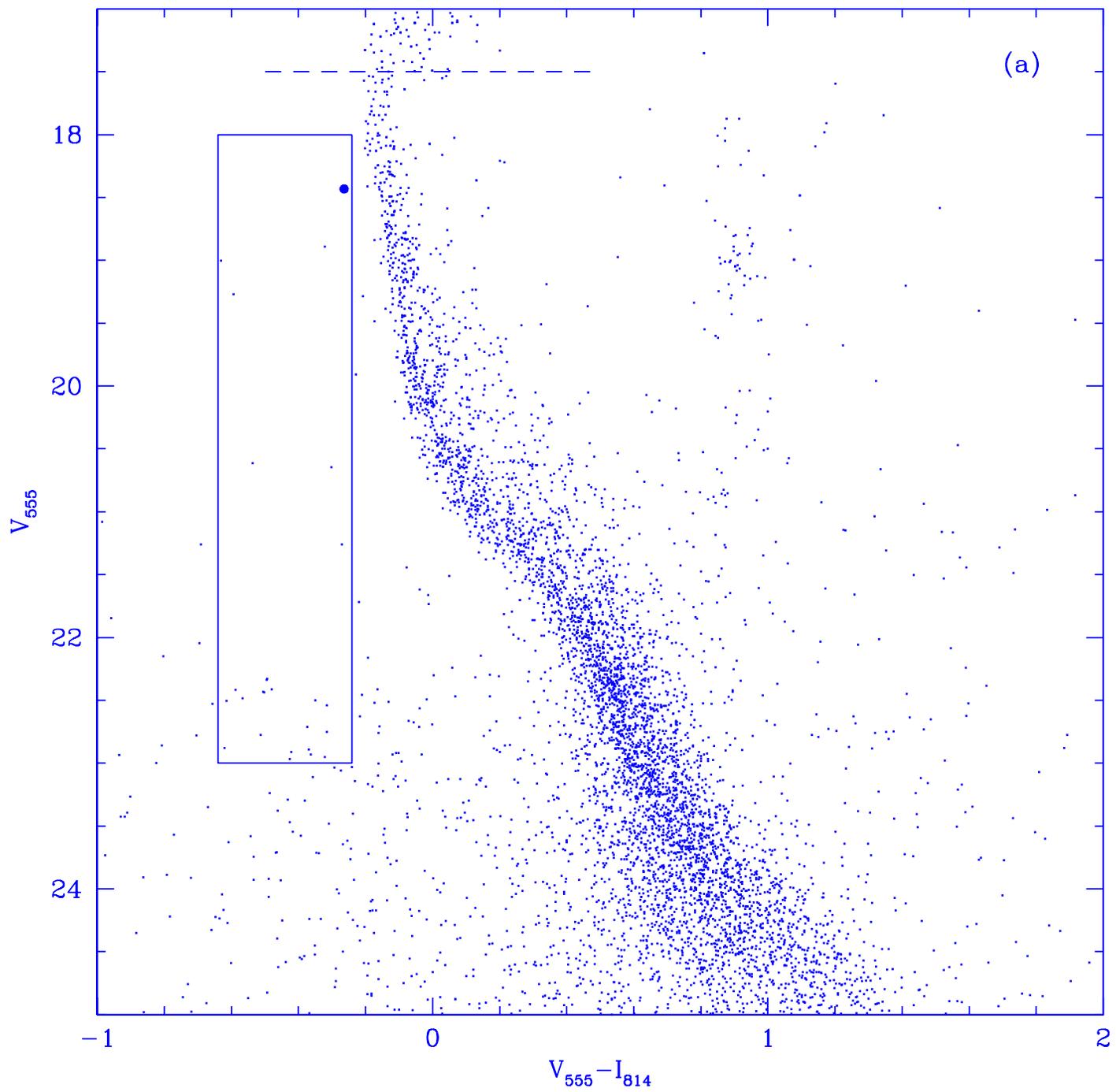
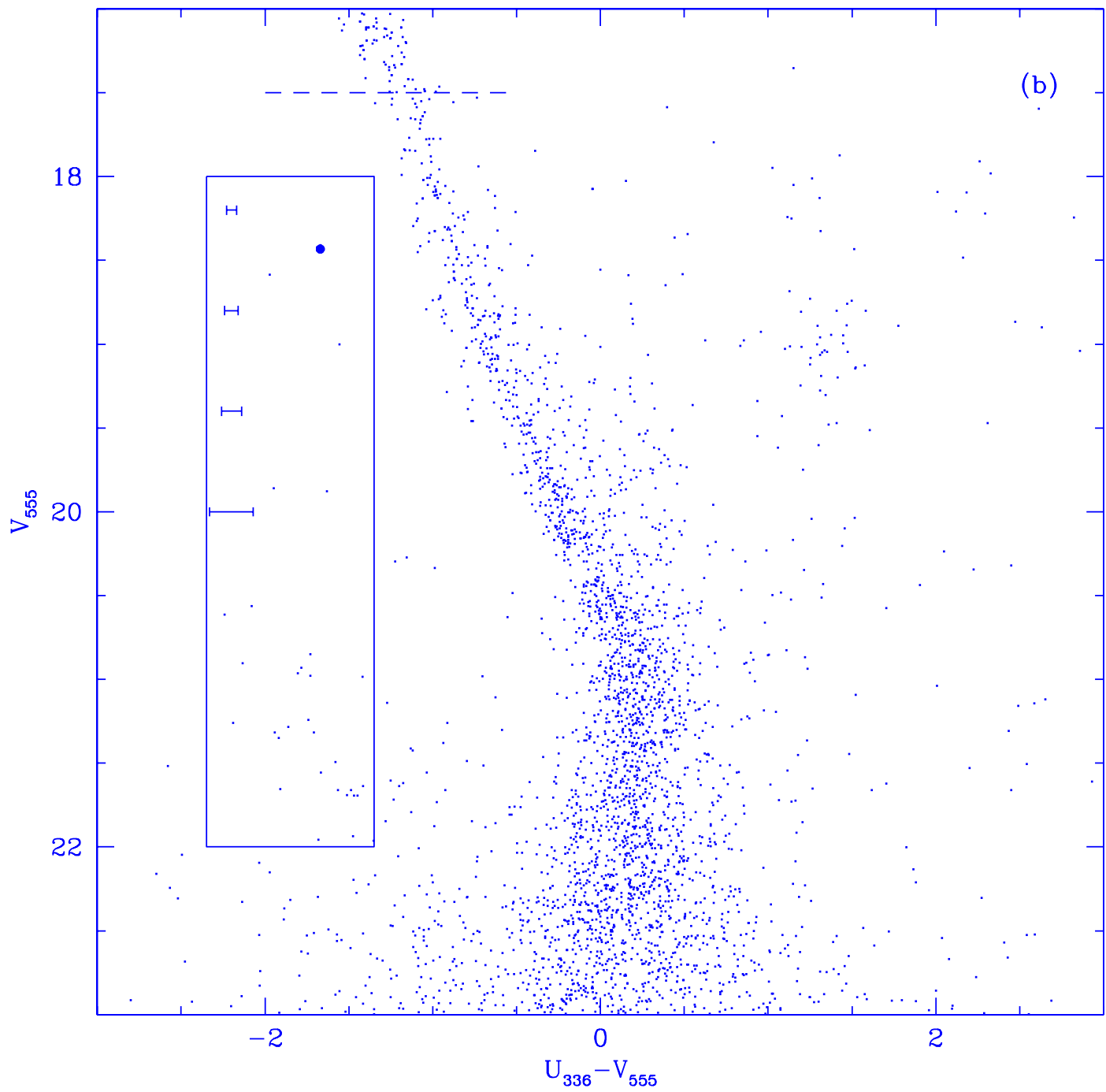
\end{document}